\documentclass[doublecol]{epl2} 

\title{Characteristics of oxygen isotope substitutions in the
quasiparticle spectrum of \chem{Bi_2Sr_2CaCu_2O_{8+\delta}}}
\shorttitle{Isotope substitution}

\author{E. Schachinger\inst{1} \and J. P. Carbotte\inst{2,3}
 \and T. Timusk\inst{2,3}}

\institute{                    
  \inst{1} Institute of Theoretical and Computational Physics -
Graz University of Technology, A-8010 Graz, Austria\\
  \inst{2} Department of Physics and Astronomy, McMaster University - Hamilton,
Ontario N1G 2W1, Canada\\
  \inst{3} The Canadian Institute for Advanced Research - Toronto, Ontario M5G
1Z8, Canada
}

\abstract{
There is an ongoing debate about the nature of the bosonic excitations
responsible for the quasiparticle self energy in high temperature
superconductors -- are they phonons or spin fluctuations? We present
a careful analysis of the bosonic excitations as revealed by the\ `kink'
feature at 70 meV in angle resolved photoemission data using Eliashberg
theory for a $d$-wave superconductor. Starting
from the assumption that nodal quasiparticles are not coupled to
the $(\pi,\pi)$ magnetic resonance, the sharp structure at $70\,$meV can be
assigned to phonons. We find that not only can we account for
the shifts of the kink energy seen on oxygen isotope substitution
but also get a quantitative estimate of the fraction of the area
under the electron-boson spectral density which is due to phonons.
We conclude that for optimally doped \chem{Bi_2Sr_2CaCu_2O_{8+\delta}}
phonons contribute $\sim 10$\% and non-phononic excitations $\sim 90$\%.
}

\begin{document}

\maketitle

%
%

Experimental evidence based on the analysis of the optical properties of
the high $T_c$ cuprate superconductors suggested early on  that the charge
carriers were strongly coupled to a spectrum of bosonic excitations in the
40 to $60\,$meV range \cite{thomas88,carbotte99}.  This was
confirmed by the discovery of a slope change or kink in electronic
dispersion of the quasiparticles (QP) by angle resolved photoemission 
spectroscopy (ARPES) in the same energy range \cite{kaminski00,bogdanov00,
johnson01}. The boson coupling that  manifests itself as peaks in the real
part of the QP self energy extracted from
the ARPES dispersion curves  can be seen in optical spectroscopy
as features in the optical scattering rate, a term in the generalized Drude
formula for
the optical conductivity \cite{thomas88,carbotte99,hwang04,carbotte05,
schachinger08a,hwang07a,hwang08,schachinger08b} and in
tunneling \cite{lee06,pasupathy08}.
Surprisingly, the nature of the bosons involved remains highly controversial
\cite{gromko03,sato03,borisenko03,hwang04,cuk04,zhou05,carbotte05,mevasana06,
kordyuk06,terashima06,
valla07,hwang07a,zhang08,schachinger08a,
hwang08,schachinger08b}. The subject is of great interest since such a
coupling, would give us a working model of superconductivity in these
materials, at least within the Migdal-Eliashberg formalism with retarded
interactions. In support of
this approach, recent numerical solutions of the Mott-Hubbard model suggest
that in the high $T_c$ cuprates retarded  interactions do provide most of the
pairing \cite{maier08,kyung08}, with the energy scale set by the
size of the antiferromagnetic exchange constant $J$. These calculations
do not support the suggestion by Anderson\cite{anderson07} that the
energy scale for the retardation might be much higher and set by the
Hubbard $U$ which would imply non-retarded pairing. Another mechanism
considered in the literature is kinetic energy pairing for which the
kinetic rather than the potential energy is lowered as the temperature
is lowered below $T_c$. In an Eliashberg formalism this can be
simulated by a phenomenological reduction in quasiparticle scattering
which could be due to a reduction in the electron-boson spectral
density \cite{schach05,carbotte06}. But this, presumably, applies only to the
superconducting state and would give no dispersion `kink' above $T_c$.

The controversy over the bosonic spectrum centers on the nature of the
bosons: are they phonons or magnetic excitations.  Neutron spectroscopy
does not resolve this problem as it shows
both phonon and magnetic excitations  in the energy region of interest.
What is needed is a fingerprint experiment, one that unambiguously
identifies the suspect bosons. In the past a number of candidate
experiments have been proposed:  doping and temperature evolution
of the bosonic spectrum \cite{johnson01,hwang04,mevasana06,kordyuk06},
substitution
with magnetic and non magnetic impurities \cite{terashima06}, but the gold
standard for the resolution of this problem is the oxygen isotope effect.
A spectral feature caused by a lattice vibration involving oxygen will
shift its frequency in a typical oxide by 6.5\% on
$^{16}\textrm{O}\to\,^{18}\textrm{O}$ substitution whereas a magnetic
excitation will
yield little or no change \cite{pailhes05}.  

There have been many studies of the effect of oxygen isotope substitutions
on the boson structure in the cuprates.  While the early ARPES experiments
could not resolve the expected $4.4\,$meV frequency shift in the kink,
an optical study
by Wang {\it et al.} \cite{wang02} on the frequency of the shoulder in
the reflectance of underdoped \chem{YBa_2Cu_3O_{7-\delta}}
found no evidence of an isotope effect.
Recent scanning tunneling microscope (STM) work has shown a shift in
the $52\,$meV structure of about the
expected amount for an oxygen mode \cite{lee06}, but this may be explained
in terms of inelastic tunneling through a barrier
\cite{pilgram06,scalapino06,hwang07b}.  Very recently new high precision
low-energy ARPES data of Iwasawa {\it et al.} ~\cite{iwasawa08} on a high
quality single crystal
of optimally doped \chem{Bi_2Sr_2CaCu_2O_{8+\delta}} (Bi2212)
with a $T_c$ of 92 K have revealed a shift of
the kink at $69\,$meV seen in the dispersion curves in the nodal direction
for $^{16}\textrm{O}\to\,^{18}\textrm{O}$ substitution with little change
in the spectrum outside
the immediate kink region. By providing a more detailed analysis of the
spectra in Ref.~\cite{iwasawa08} we will show here that not only do these
new measurements provide a fingerprint for the presence of a phonon
contribution to the self energy of the charge carriers, but that they permit
us to calculate the fraction of the bosonic coupling parameter
$\lambda$ associated with phonons and the fraction with magnetic excitations
and conclude that the phonon contribution is $\sim10$\% of the area
under the spectral density. The major contribution to the glue is
non phononic. 

A key to an understanding of the bosonic structure in ARPES is the
recognition that there are several features that can contribute to the
kink in the dispersion curves and that the challenge is to separate
their contributions. We will show that the new experiments of
Iwasawa {\it et al.} \cite{iwasawa08} allow us to do that unambiguously.
Calculations of
scattering by spin fluctuations suggest that an important role is played
by bosons that have a $(\pi,\pi)$ wave vector and they are particularly
efficient in scattering charge carriers near the antinodes but have
little effect on the nodal quasiparticles. Neutron scattering experiments
show that the local ({\bf q} averaged) susceptibility has a strong peak in
energy around $40\,$meV, but also that the {\bf q} dependent susceptibility
is peaked at $(\pi,\pi)$ in this energy range. As a result, in frequency
dependent measurements, the interaction with  this magnetic peak
is small in the nodal direction because Fermi surface to
Fermi surface electronic transitions involving the spin one resonance
with momentum transfer of $(\pi,\pi)$ cannot take place due to the
particular Fermi surface geometry but this becomes more possible as
we move towards the antinodal direction.
This agrees with ARPES measurements that show the growth of
a sharp feature as the momentum of the quasiparticles approaches the
zone boundary at $(0,\pi)$~\cite{cuk04}.  

What then scatters the quasiparticles in the nodal direction?
We suggest that spin excitations that are not centered
at $(\pi,\pi)$ but spread out as a continuum and phonons that are also
expected to be scattered more or less independent of QP vector {\bf q}.
ARPES measurements \cite{zhang08} show that while there is a sharp
spectral feature
in the $65\,$meV region for nodal quasiparticles it is weak.
Nevertheless, as we have argued above, it cannot be assigned
to the $(\pi,\pi)$ spin resonance and we suggest this feature
to be the residual phonon scattering superimposed
on a broad magnetic background. In what follows we will show that this
idea is fully consistent with all the data of Iwasawa {\it et al.}
\cite{iwasawa08} and
thus provides us with a quantitative tool to estimate the relative
contribution of phonons and spin fluctuations to high temperature
superconductivity. 

\begin{figure}[pt]
  \onefigure{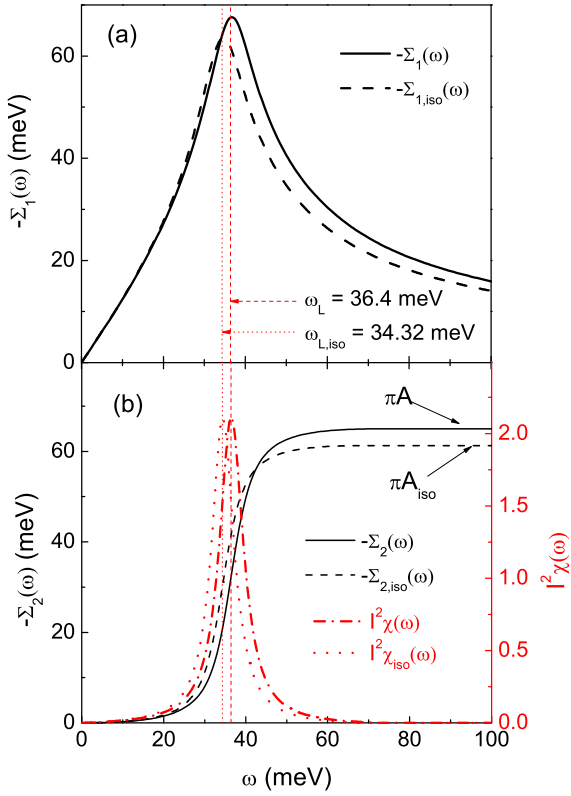}
  \caption{(Colour online)
(a) Minus the real part of the QP self energy, $-\Sigma_1(\omega)$
($^{16}$O solid and $^{18}$O dashed lines) vs energy $\omega$.
The two thin dashed and dotted vertical
lines indicate the position of the peak in the $I^2\chi(\omega)$ spectrum,
$\omega_\textrm{L}$ and $\omega_\textrm{L,iso}$, respectively. (b) The same as
(a) but now for $-\Sigma_2(\omega)$. Lorentzian model
$I^2\chi(\omega)$ spectrum for $^{16}$O (dashed-dotted
line) and $I^2\chi_\textrm{iso}(\omega)$
for $^{18}$O (dotted line) right hand scale.
}
  \label{fig:1}
\end{figure}

The changes in the real and imaginary parts of the self energy with
isotopic substitution are best illustrated with a simple model for
the effect of a single Einstein mode at frequency $\omega_E$ on the
self energy. To simplify the algebra we first show the results for a
mode of zero width. The electron-boson interaction spectral density
$I^2\chi(\omega) = A\delta(\omega-\omega_E)$ where $A$ is the area
under the Dirac
$\delta$-function $\delta(x)$. The resulting QP self energy
in the normal state at zero temperature $(T=0)$ is
\begin{equation}
  \label{eq:1}
  \Sigma(\omega) = \Sigma_1(\omega)+i\Sigma_2(\omega)
    = A\ln\left\vert\frac{\omega_E-\omega}{\omega_E+\omega}
           \right\vert-i\pi A\theta(\vert\omega\vert-\omega_E),
\end{equation}
where $\theta(x)$ is a step function. The dotted and the dash-dotted curves
in Fig.~\ref{fig:1}(b) illustrate these functions where we have modified
the spectral
density by giving it a Lorentzian width. We note that under isotopic
substitution $^{16}$O $\to$ $^{18}$O, the frequency of the mode is reduced
to $\omega_E \to \gamma\omega_E$, but the coupling constant too is reduced
to $A \to \gamma A = A_\textrm{iso}$ where $\gamma = \sqrt{16/18} = 0.94$.
The mass enhancement
parameter $\lambda = 2A/\omega_E$ remains unchanged. In the limit
$\omega\ll \omega_E$ the real part of the self energy is linear
in $\omega$ with slope $\lambda$. As Fig.~\ref{fig:1} shows, isotopic
substitution leaves two 'fingerprints`  in the self energy spectra,
the well known shift of the mode frequency by 6\% but a second, less known
one, a reduction in amplitude of 6\%.  In the real part of the self energy,
Fig.~\ref{fig:1}(a),
the peak is shifted to lower energy and reduced in amplitude by the same
factor while for the imaginary part, Fig.~\ref{fig:1}(b),
the onset of scattering is also moved
to lower energies and its saturation value is reduced by the same $\gamma$.
The inflection points are found at $\omega_L$
and $\omega_{L,iso} = \gamma\omega_L$ respectively.

Many of the  features of the model
of a single bosonic mode shown in Fig.~\ref{fig:1} do not agree with recent
data of Iwasawa {\it et al.} \cite{iwasawa08} even when we account for
the changes in QP properties that are brought about by superconductivity.
 First, coupling,
predominantly to a single mode peaked around $\omega_L$, whatever its
origin, leads to a saturation in the magnitude of the QP
scattering rate, $\Sigma_2(\omega)$, at frequencies above $\omega_L$.
Such a saturation is not seen in the published ARPES data obtained
by fitting a Lorentzian form to momentum distribution curves,
nor is it seen in optical data. Both methods find scattering rates
\cite{valla07,hwang07a} which continue to increase up to energies of
order $400\,$meV. This is well beyond any phonon energy
($\sim 86\,$meV in Bi2212 \cite{renker89}) and shows that in addition
to the sharp bosonic peak
the charge carriers are also coupled to high energy boson modes of some other
origin. Secondly, the model predicts a reduction of the amplitude of the real
part of the self energy of 6\% but the data of Iwasawa {\it et al.}
shows little change in the range
around $70\,$meV. While the {\it frequency} of the peak is shifted by the
expected  amount, the magnitude of the maximum in $-\Sigma_1(\omega)$
appears  not to be significantly changed. This contradicts the single
phonon model of the self energy. 

We will next show that these contradictions can be resolved by assuming
that in addition to a (weak) phonon contribution there is a strong
contribution from magnetic scattering and that there is additional
influence due to finite band widths. Taking this into account
we are able to estimate the
relative strength of the phonon and magnetic contributions to the
QP self energy.


\begin{figure}[pt]
  \vspace*{-7mm}
  \centerline{\includegraphics[width=10cm]{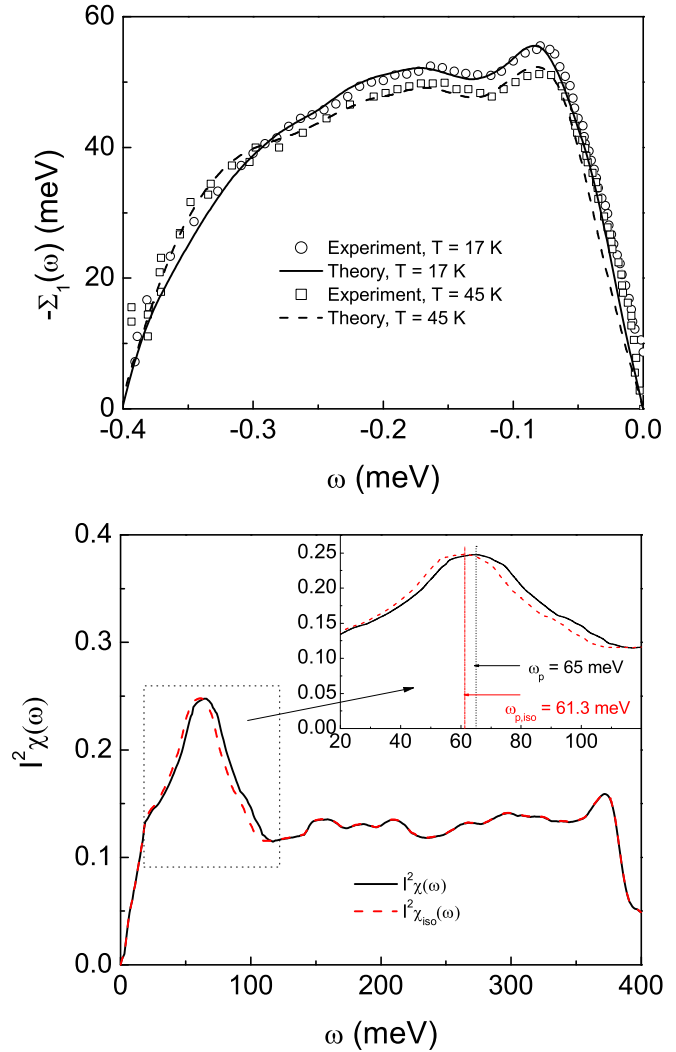}}
  \caption{(Colour online)
(a)  Presents fits to the real part of the superconducting
state QP self energy ARPES data of Ref. \protect{\cite{zhang08}}
as reported by Schachinger and Carbotte \cite{schachinger08a}
for $T=17\,$K (solid line and open diamonds) and $45\,$K
(dashed line and open squares).
(b) The $I^2\chi(\omega)$ spectrum from maximum entropy inversion
of the $T=17\,$K Zhang {\it et al.} \protect{\cite{zhang08}}
data (solid line) with $\lambda = 1.19$, and
the area under the spectrum $A = 51.9\,$meV. The (red) dashed line
shows the spectrum $I^2\chi_\textrm{iso}(\omega)$ which was used
to simulate the isotope effect (depicted also in the inset for
clarity). Only the area of the peak equal
to $6.15\,$meV or a partial $\lambda$ of 0.21
is shifted in energy by 6\% with $\lambda = 1.19$ and
$A_\textrm{iso} = 51.4\,$meV.
}
  \label{fig:2}
\end{figure}

To accomplish this we use the recent high resolution ARPES work of
Zhang {\it et al.} \cite{zhang08}. These data were analyzed using maximum
entropy inversion \cite{schachinger08a} in the superconducting state
based on finite band width $d$-wave Eliashberg equations, in order to
determine
the spectral density $I^2\chi(\omega)$ of the bosons contributing to
the nodal direction QP self energy. In Fig.~\ref{fig:2}(a) we show our
fits, solid curve for $T=17\,$K (dashed curve for $T=45\,$K) to the
ARPES data, open circles (open squares).
Results for the electron-boson spectral density $I^2\chi(\omega)$
at $T=17\,$K and a bandwidth of $1.2\,$eV are shown
in Fig.~\ref{fig:2}(b) as the solid (black) line. This
$I^2\chi(\omega)$ spectrum is very different
from the single sharp Lorentzian form used so far. It has
a broad peak around $\omega_p = 65\,$meV superimposed on
a large background extending to $400\,$meV. The cutoff
in this spectrum depends critically on the choice of the bare dispersion
curve made in the ARPES study. The renormalizations end at the crossing
between bare and
dressed dispersions and in the work of Zhang {\it et al.} \cite{zhang08}
this was taken to be
$400\,$meV. This choice is consistent with optical data \cite{
hwang07a,hwang08,schachinger08b,carbotte05}.
It is also consistent with the observation that the QP as
well as optical scattering rates still increase
with increasing $\omega$ in the range of a few $100\,$meV.

\begin{figure}[pt]
\vspace*{-5mm}
\onefigure{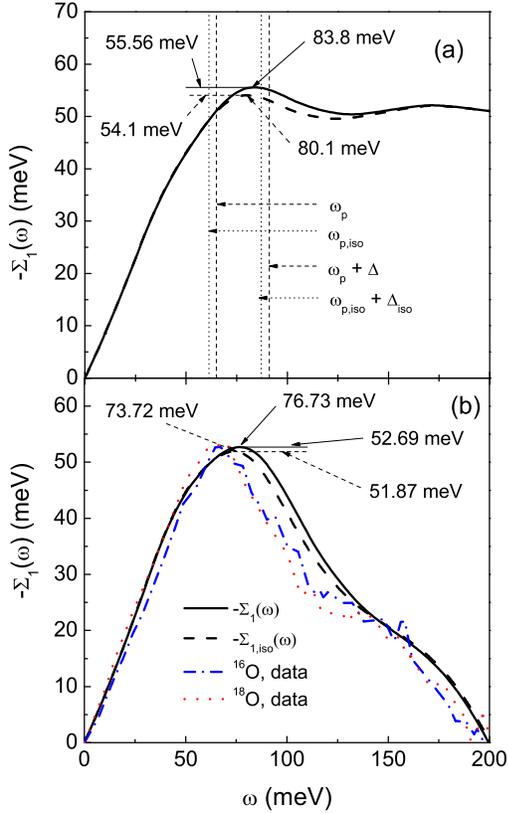}
\caption{(Colour online)
 The real part of the QP self energy
$-\Sigma_1(\omega)$ vs energy $\omega$ of Bi2212 in the superconducting state
at $T=17\,$K. The solid line corresponds to the solid line
in the inset of Fig.~\ref{fig:2}. The dashed line is the result due
to the isotope effect simulated by the spectral function
$I^2\chi_\textrm{iso}(\omega)$. The peak position shifts down
by $4.5$\% and its amplitude by 3\%.
(b) The same as (a) but for
a cutoff of $200\,$meV in the spectra of Fig.~\ref{fig:2}.
The dashed-dotted and dotted lines correspond to
the $^{16}$O and $^{18}$O data for cut zero presented
by Iwasawa {\it et al.} \cite{iwasawa08} in their Fig.~2(b) rescaled
to meet our theoretical results in amplitude.
}
  \label{fig:3}
\end{figure}

Our model of the isotope substitution on the real part
of the self energy is based on
{\em shifting only the peak} in $I^2\chi(\omega)$
shown by the (red) dashed curve in Fig.~\ref{fig:2}(b) leaving the
background unchanged as depicted in the inset, and results for the
self energy
are shown in Fig.~\ref{fig:3}(a). The solid curve is
for $^{16}$O and the dashed curve for $^{18}$O.
The results are in good overall agreement with the experimental
data of Iwasawa {\it et al.} \cite{iwasawa08}. For instance,
we predict a shift in the position of the peak
of the order of 4.5\% against the observation of
$5\pm0.8\,$\%. We also predict a reduction in
the amplitude of the self energy maximum of $\sim\!3$\% somewhat larger
than is observed in the data.
Note that the position of the peak in $-\Sigma_1(\omega)$ is not
trivially correlated to the position of the peak in the spectrum,
$\omega_p$ ($\omega_{p,iso}$), nor to their values
displaced by the gap amplitude, $\omega_p+\Delta$
($\omega_{p,iso}+\Delta_{iso}$).

Figure~\ref{fig:3}(b) presents the results of a numerical
simulation using a smaller energy cutoff of  $200\,$meV
as done by Iwasawa {\it et al.} \cite{iwasawa08}.
The solid and dashed lines are for $^{16}$O and
$^{18}$O, respectively. The temperature is $T=17\,$K and the bandwidth
had to be reduced to $0.6\,$eV.
We also added experimental data for $^{18}$O
(dashed-dotted line) and for $^{16}$O (dotted line) from
Fig.~2(b) (cut zero) of Iwasawa {\it et al.} rescaled to meet
our $\Sigma_1(\omega)$ data in amplitude. The qualitative agreement between
data and numerical simulation is now even better. As a result of the
new cutoff the peak position in our theoretical results moved down by
$\sim 6\,$meV.
Also the difference in amplitude is now less than one meV and
this explains why almost no difference in amplitude can
be seen in the data.
As a result of this simulation we can safely conclude that the data
can be reproduced by a
$I^2\chi(\omega)$ spectrum which has the same shape as the
spectrum presented by the solid line in
Fig.~\ref{fig:2} with the broad peak which will be shifted by
isotope substitution now centered between 55 and
$58\,$meV.

The area under the electron-boson spectral density which we have shifted
in energy by 6\% corresponds to $6.13\,$meV or about 10\% of the total
area under the $I^2\chi(\omega)$.  If we attribute this shift in
area entirely to an electron-phonon coupling, the mass enhancement
factor involved is $\lambda\simeq 0.2$. This value is much smaller
than estimated in Ref.~\cite{iwasawa08} but is of the order found in
band structure calculations on related cuprates
\cite{savrasov96,bohnen03,giustino08}.
It also compares favorably with the total value of mass enhancement
$\lambda_{tot}=0.23$ found in studies \cite{deveraux04} of anisotropic
electron-phonon coupling due to the oxygen buckling mode ($\sim 36\,$meV).
The coupling to the breathing mode ($\sim 70\,$meV) was found to be much
smaller: $\lambda_{tot} \sim 0.02$. Our value of $\lambda = 0.2$ corresponds
to less than 20\% of the total $\lambda$ found through maximum entropy
inversion \cite{schachinger08a} of the nodal direction ARPES data
of Ref.~\cite{zhang08}. 

In summary we analyzed recent nodal direction ARPES data of the effect
of $^{16}\textrm{O}\to\,^{18}\textrm{O}$ substitutions on the QP
self energy. We find that the data are consistent with a model where only
$\sim10$\% of the self energy is through coupling to lattice vibrations,
so that the major part of the glue function is of another origin, we
believe spin fluctuations. 
We conclude that the new measurements need not imply that phonons
play a large role in the superconductivity of these materials.
If the mass enhancement parameter $\lambda$ is used instead of the
area under $I^2\chi(\omega)$ as a measure of the strength of the
electronic renormalizations, the phonon contribution to $\lambda$ is
0.2 or less than 20\% of the total $\lambda = 1.19$ for our realistic
model for the spectral density in \chem{Bi_2Sr_2CaCu_2O_{8+\delta}}
obtained from inversion of nodal direction ARPES data.

\acknowledgments
Research supported in part by the Natural Sciences and Engineering Research
Council of Canada (NSERC) and by the Canadian Institute for Advanced Research
(CIFAR).

\end{document}